%% file: main.tex
\pdfoutput=1

\documentclass[12pt,a4paper]{article}

\usepackage{ifthen} 
\newboolean{pdflatex}
\setboolean{pdflatex}{true} 

\newboolean{articletitles}
\setboolean{articletitles}{true} 

\newboolean{uprightparticles}
\setboolean{uprightparticles}{false} 

\newboolean{inbibliography}
\setboolean{inbibliography}{false} 

\input{preamble}
\usepackage{longtable} 

\begin{document}

\renewcommand{\thefootnote}{\fnsymbol{footnote}}
\setcounter{footnote}{1}

\input{title-LHCb-PAPER}

\renewcommand{\thefootnote}{\arabic{footnote}}
\setcounter{footnote}{0}



\pagestyle{plain} 
\setcounter{page}{1}
\pagenumbering{arabic}



\input{introduction}

\input{detector}

\input{selection}

\input{fits}

\input{systematics}

\input{conclusion}

\input{acknowledgements}

\addcontentsline{toc}{section}{References}
\setboolean{inbibliography}{true}
\bibliographystyle{LHCb}
\bibliography{main,LHCb-PAPER,LHCb-CONF,LHCb-DP,LHCb-TDR}

\newpage

\newpage
\input{LHCb_Authorship_flat_20-Sep-2016.tex}

\end{document}

%% file: preamble.tex
\usepackage[top=1in, bottom=1.25in, left=1in, right=1in]{geometry}

\usepackage{microtype}
\usepackage{lineno}  
\usepackage{xspace} 
\usepackage{caption} 

\usepackage{graphicx}  
\usepackage{color}
\usepackage{colortbl}
\graphicspath{{./figs/}} 

\usepackage{amsmath} 
\usepackage{amssymb}
\usepackage{amsfonts}
\usepackage{upgreek} 

\newcommand*\patchAmsMathEnvironmentForLineno[1]{%
\expandafter\let\csname old#1\expandafter\endcsname\csname #1\endcsname
\expandafter\let\csname oldend#1\expandafter\endcsname\csname
end#1\endcsname
 \renewenvironment{#1}%
   {\linenomath\csname old#1\endcsname}%
   {\csname oldend#1\endcsname\endlinenomath}%
}
\newcommand*\patchBothAmsMathEnvironmentsForLineno[1]{%
  \patchAmsMathEnvironmentForLineno{#1}%
  \patchAmsMathEnvironmentForLineno{#1*}%
}
\AtBeginDocument{%
\patchBothAmsMathEnvironmentsForLineno{equation}%
\patchBothAmsMathEnvironmentsForLineno{align}%
\patchBothAmsMathEnvironmentsForLineno{flalign}%
\patchBothAmsMathEnvironmentsForLineno{alignat}%
\patchBothAmsMathEnvironmentsForLineno{gather}%
\patchBothAmsMathEnvironmentsForLineno{multline}%
\patchBothAmsMathEnvironmentsForLineno{eqnarray}%
}

\usepackage{hyperref}    
\usepackage[all]{hypcap} 

\input{lhcb-symbols-def} 

\input{our-defs} 

\usepackage{cite} 
\usepackage{mciteplus}

%% file: lhcb-symbols-def.tex
\usepackage{xspace} 
\usepackage{upgreek}

\def\lhcb {\mbox{LHCb}\xspace}

\def\babar  {\mbox{BaBar}\xspace}
\def\belle  {\mbox{Belle}\xspace}
\def\cleo   {\mbox{CLEO}\xspace}

\def\MagUp {\mbox{\em Mag\kern -0.05em Up}\xspace}

\ifthenelse{\boolean{uprightparticles}}%
{

 \def\Ppi         {\ensuremath{\uppi}\xspace}

 \def\PDelta      {\ensuremath{\Delta}\xspace}                 
 \def\PXi      {\ensuremath{\Xi}\xspace}                 
 \def\PLambda      {\ensuremath{\Lambda}\xspace}                 
 \def\PSigma      {\ensuremath{\Sigma}\xspace}                 
 \def\POmega      {\ensuremath{\Omega}\xspace}                 
 \def\PUpsilon      {\ensuremath{\Upsilon}\xspace}                 
 
 \def\PB      {\ensuremath{\mathrm{B}}\xspace}                 
                  
 \def\PD      {\ensuremath{\mathrm{D}}\xspace}

 \def\PK      {\ensuremath{\mathrm{K}}\xspace}

 \def\Pb      {\ensuremath{\mathrm{b}}\xspace}                 
 \def\Pc      {\ensuremath{\mathrm{c}}\xspace}                 
                  
 \def\Pe      {\ensuremath{\mathrm{e}}\xspace}

 \def\Pi      {\ensuremath{\mathrm{i}}\xspace}

 \def\Pp      {\ensuremath{\mathrm{p}}\xspace}

 \def\Ps      {\ensuremath{\mathrm{s}}\xspace}                 
                  
 \def\Pu      {\ensuremath{\mathrm{u}}\xspace}

}
{

 \def\Ppi         {\ensuremath{\pi}\xspace}

 \mathchardef\PDelta="7101
 \mathchardef\PXi="7104
 \mathchardef\PLambda="7103
 \mathchardef\PSigma="7106
 \mathchardef\POmega="710A
 \mathchardef\PUpsilon="7107
                  
 \def\PB      {\ensuremath{B}\xspace}                 
                  
 \def\PD      {\ensuremath{D}\xspace}

 \def\PK      {\ensuremath{K}\xspace}

 \def\Pb      {\ensuremath{b}\xspace}                 
 \def\Pc      {\ensuremath{c}\xspace}                 
                  
 \def\Pe      {\ensuremath{e}\xspace}

 \def\Pi      {\ensuremath{i}\xspace}

 \def\Pp      {\ensuremath{p}\xspace}

 \def\Ps      {\ensuremath{s}\xspace}                 
                  
 \def\Pu      {\ensuremath{u}\xspace}

}

\makeatletter
\ifcase \@ptsize \relax
  \newcommand{\miniscule}{\@setfontsize\miniscule{4}{5}}
\or
  \newcommand{\miniscule}{\@setfontsize\miniscule{5}{6}}
\or
  \newcommand{\miniscule}{\@setfontsize\miniscule{5}{6}}
\fi
\makeatother

\DeclareRobustCommand{\optbar}[1]{\shortstack{{\miniscule (\rule[.5ex]{1.25em}{.18mm})}
  \\ [-.7ex] $#1$}}

\def\epem       {{\ensuremath{\Pe^+\Pe^-}}\xspace}

\def\uquark    {{\ensuremath{\Pu}}\xspace}

\def\squark    {{\ensuremath{\Ps}}\xspace}

\def\cquark    {{\ensuremath{\Pc}}\xspace}

\def\bquark    {{\ensuremath{\Pb}}\xspace}
\def\bquarkbar {{\ensuremath{\overline \bquark}}\xspace}

\def\pion   {{\ensuremath{\Ppi}}\xspace}

\def\pip    {{\ensuremath{\pion^+}}\xspace}
\def\pim    {{\ensuremath{\pion^-}}\xspace}

\def\pimp   {{\ensuremath{\pion^\mp}}\xspace}

\def\kaon    {{\ensuremath{\PK}}\xspace}
  \def\Kbar    {{\kern 0.2em\overline{\kern -0.2em \PK}{}}\xspace}

\def\KorKbar    {\kern 0.18em\optbar{\kern -0.18em K}{}\xspace}
\def\Kz      {{\ensuremath{\kaon^0}}\xspace}

\def\Kp      {{\ensuremath{\kaon^+}}\xspace}
\def\Km      {{\ensuremath{\kaon^-}}\xspace}
\def\Kpm     {{\ensuremath{\kaon^\pm}}\xspace}

\def\KS      {{\ensuremath{\kaon^0_{\mathrm{ \scriptscriptstyle S}}}}\xspace}

  \def\Dbar    {{\kern 0.2em\overline{\kern -0.2em \PD}{}}\xspace}
\def\D       {{\ensuremath{\PD}}\xspace}

\def\DorDbar    {\kern 0.18em\optbar{\kern -0.18em D}{}\xspace}
\def\Dz      {{\ensuremath{\D^0}}\xspace}

\def\Ds      {{\ensuremath{\D^+_\squark}}\xspace}

\def\B       {{\ensuremath{\PB}}\xspace}
\def\Bbar    {{\ensuremath{\kern 0.18em\overline{\kern -0.18em \PB}{}}}\xspace}

\def\BorBbar    {\kern 0.18em\optbar{\kern -0.18em B}{}\xspace}

\def\Bu      {{\ensuremath{\B^+}}\xspace}

\def\Bd      {{\ensuremath{\B^0}}\xspace}
\def\Bs      {{\ensuremath{\B^0_\squark}}\xspace}

  \def\Y#1S{\ensuremath{\PUpsilon{(#1S)}}\xspace}

\def\proton      {{\ensuremath{\Pp}}\xspace}
\def\antiproton  {{\ensuremath{\overline \proton}}\xspace}

\def\Lz          {{\ensuremath{\PLambda}}\xspace}
\def\Lbar        {{\ensuremath{\kern 0.1em\overline{\kern -0.1em\PLambda}}}\xspace}
\def\LorLbar    {\kern 0.18em\optbar{\kern -0.18em \PLambda}{}\xspace}

\def\Lc      {{\ensuremath{\Lz^+_\cquark}}\xspace}

\def\BF         {{\ensuremath{\mathcal{B}}}\xspace}

\newcommand{\decay}[2]{\ensuremath{#1\!\to #2}\xspace}         

\def\to                 {\ensuremath{\rightarrow}\xspace}

\def\Vub  {{\ensuremath{V_{\uquark\bquark}}}\xspace}

\def\AT#1     {\ensuremath{A_{\mathrm{T}}^{#1}}\xspace}           

\def\C#1      {\ensuremath{\mathcal{C}_{#1}}\xspace}                       
\def\Cp#1     {\ensuremath{\mathcal{C}_{#1}^{'}}\xspace}                    
\def\Ceff#1   {\ensuremath{\mathcal{C}_{#1}^{\mathrm{(eff)}}}\xspace}        
\def\Cpeff#1  {\ensuremath{\mathcal{C}_{#1}^{'\mathrm{(eff)}}}\xspace}       
\def\Ope#1    {\ensuremath{\mathcal{O}_{#1}}\xspace}                       
\def\Opep#1   {\ensuremath{\mathcal{O}_{#1}^{'}}\xspace}                    

\newcommand{\tev}{\ifthenelse{\boolean{inbibliography}}{\ensuremath{~T\kern -0.05em eV}\xspace}{\ensuremath{\mathrm{\,Te\kern -0.1em V}}}\xspace}
\newcommand{\gev}{\ensuremath{\mathrm{\,Ge\kern -0.1em V}}\xspace}
\newcommand{\mev}{\ensuremath{\mathrm{\,Me\kern -0.1em V}}\xspace}
\newcommand{\kev}{\ensuremath{\mathrm{\,ke\kern -0.1em V}}\xspace}
\newcommand{\ev}{\ensuremath{\mathrm{\,e\kern -0.1em V}}\xspace}
\newcommand{\gevc}{\ensuremath{{\mathrm{\,Ge\kern -0.1em V\!/}c}}\xspace}
\newcommand{\mevc}{\ensuremath{{\mathrm{\,Me\kern -0.1em V\!/}c}}\xspace}
\newcommand{\gevcc}{\ensuremath{{\mathrm{\,Ge\kern -0.1em V\!/}c^2}}\xspace}
\newcommand{\gevgevcccc}{\ensuremath{{\mathrm{\,Ge\kern -0.1em V^2\!/}c^4}}\xspace}
\newcommand{\mevcc}{\ensuremath{{\mathrm{\,Me\kern -0.1em V\!/}c^2}}\xspace}

\def\mum  {\ensuremath{{\,\upmu\mathrm{m}}}\xspace}

\def\invfb   {\ensuremath{\mbox{\,fb}^{-1}}\xspace}

\newcommand{\chisq}{\ensuremath{\chi^2}\xspace}

\newcommand{\chisqip}{\ensuremath{\chi^2_{\text{IP}}}\xspace}

\def\gsim{{~\raise.15em\hbox{$>$}\kern-.85em
          \lower.35em\hbox{$\sim$}~}\xspace}
\def\lsim{{~\raise.15em\hbox{$<$}\kern-.85em
          \lower.35em\hbox{$\sim$}~}\xspace}

\def\ptot       {\mbox{$p$}\xspace}
\def\pt         {\mbox{$p_{\mathrm{ T}}$}\xspace}

\def\evtgen     {\mbox{\textsc{EvtGen}}\xspace}

\def\geant      {\mbox{\textsc{Geant4}}\xspace}

\def\photos     {\mbox{\textsc{Photos}}\xspace}

\def\pythia     {\mbox{\textsc{Pythia}}\xspace}

\def\tell1  {TELL1\xspace}
\def\ukl1   {UKL1\xspace}

\newcommand{\ie}{\mbox{\itshape i.e.}\xspace}

%% file: our-defs.tex
\def\Bds{{\ensuremath{\B^0_{\kern -0.1em{\scriptscriptstyle (}\kern -0.05em\squark\kern -0.03em{\scriptscriptstyle )}}}}\xspace}

\def\Vz {\mbox{$V^0$}\xspace}

\def\Sbar{\kern 0.2em\overline{\kern -0.2em \Sigma}{}\xspace}

\newcommand{\BuPLbar}{\texorpdfstring{\decay{\Bu}{\proton \Lbar}}{}}

\newcommand{\BuKSPi}{\texorpdfstring{\decay{\Bu}{\KS \pip}}{}}
\newcommand{\BuKSK}{\texorpdfstring{\decay{\Bu}{\KS \Kp}}{}}
\newcommand{\BuKSH}{\texorpdfstring{\decay{\Bu}{\KS h^+}}{}}

\newcommand{\BPPbar}{\texorpdfstring{\decay{\Bds}{\proton \antiproton}}{}}
\newcommand{\BdPPbar}{\texorpdfstring{\decay{\Bd}{\proton \antiproton}}{}}

\newcommand{\BuPLbarExcited}{\texorpdfstring{\decay{\Bu}{\proton \Lbar(1520)}}{}}

\newcommand{\BdKSPiPi}{\texorpdfstring{\decay{\Bd}{\KS \pip \pim}}{}}
\newcommand{\BdKSKK}{\texorpdfstring{\decay{\Bd}{\KS \Kp \Km}}{}}

\newcommand{\BuPPbarPi}{\texorpdfstring{\decay{\Bu}{\proton \antiproton \pip}}{}}

\newcommand{\BsKSKPi}{\texorpdfstring{\decay{\Bs}{\KS \Kpm \pimp}}{}}
\newcommand{\BsKSPiPi}{\texorpdfstring{\decay{\Bs}{\KS \pip \pim}}{}}

\newcommand{\BKSHH}{\texorpdfstring{\decay{\Bds}{\KS h^{+} h^{'-}}}{}}

\newcommand{\LcPKPi}{\texorpdfstring{\decay{\Lc}{\proton \Km \pip}}{}}

\newcommand{\BuPSbar}{\texorpdfstring{\decay{\Bu}{\proton \Sbar^0}}{}}

\newcommand{\KSPiPi}{\texorpdfstring{\decay{\KS}{\pip \pim}}{}}
\newcommand{\LPPi}{\texorpdfstring{\decay{\PLambda}{\proton \pim}}{}}

\def\PLbar {$\proton\Lbar$\xspace}

\def\KSPi  {\ensuremath{\KS\pip}\xspace}

\def\splot{\mbox{\em sPlot}\xspace}

%% file: title-LHCb-PAPER.tex

\begin{titlepage}
\pagenumbering{roman}

\vspace*{-1.5cm}
\centerline{\large EUROPEAN ORGANIZATION FOR NUCLEAR RESEARCH (CERN)}
\vspace*{1.5cm}
\noindent
\begin{tabular*}{\linewidth}{lc@{\extracolsep{\fill}}r@{\extracolsep{0pt}}}
\ifthenelse{\boolean{pdflatex}}
{\vspace*{-2.7cm}\mbox{\!\!\!\includegraphics[width=.14\textwidth]{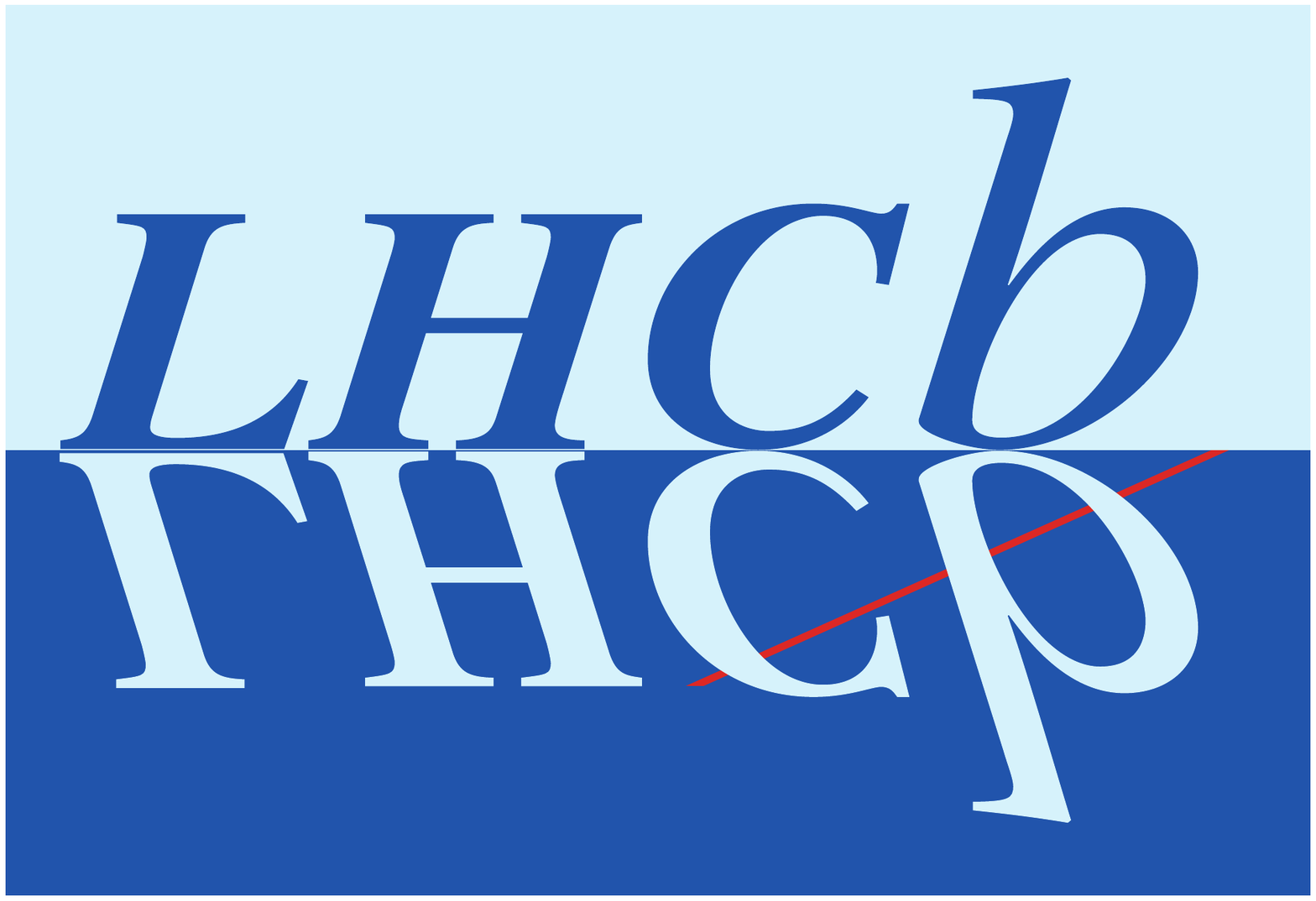}} & &}%
{\vspace*{-1.2cm}\mbox{\!\!\!\includegraphics[width=.12\textwidth]{lhcb-logo.eps}} & &}%
\\
 & & CERN-EP-2016-275 \\  
 & & LHCb-PAPER-2016-048 \\  
 & & 28 April 2017 \\ 
 & & \\
\end{tabular*}

\vspace*{4.0cm}

{\normalfont\bfseries\boldmath\huge
\begin{center}
Evidence for the two-body\\
charmless baryonic decay \BuPLbar
\end{center}
}

\vspace*{2.0cm}

\begin{center}
The LHCb collaboration\footnote{Authors are listed at the end of this paper.}
\end{center}

\vspace{\fill}

\begin{abstract}
  \noindent
A search for the rare two-body charmless baryonic decay
\BuPLbar is performed with $pp$ collision data,
corresponding to an integrated luminosity of 3~\invfb,
collected by the LHCb experiment at centre-of-mass energies of 7 and 8~\tev.
An excess of \BuPLbar candidates with respect to background expectations
is seen with a statistical significance of 4.1 standard deviations,
and constitutes the first evidence for this decay.
The branching fraction, measured using the \BuKSPi decay for normalisation, is
\begin{eqnarray}
\BF(\BuPLbar) & = & ( 2.4 \,^{+1.0}_{-0.8} \pm 0.3 ) \times 10^{-7} \,, \nonumber
\end{eqnarray}
where the first uncertainty is statistical and the second systematic.
\end{abstract}

\vspace*{2.0cm}

\begin{center}
  Published in JHEP 04 (2017) 162
\end{center}

\vspace{\fill}

{\footnotesize 
\centerline{\copyright~CERN on behalf of the \lhcb collaboration, licence \href{http://creativecommons.org/licenses/by/4.0/}{CC-BY-4.0}.}}
\vspace*{2mm}

\end{titlepage}


\newpage
\setcounter{page}{2}
\mbox{~}
%

\cleardoublepage

%% file: introduction.tex
\section{Introduction}
\label{sec:introduction}
The experimental study of $B$ meson decays to baryonic final states has a long history,
including numerous searches and observations by the asymmetric $e^+ e^-$ collider
experiments \babar and \belle~\cite{Bevan:2014iga}.
In recent years the \lhcb collaboration reported the first observation
of a two-body charmless baryonic \Bu decay and the first evidence for a similar \Bd decay,
namely \BuPLbarExcited~\cite{LHCb-PAPER-2014-034} and \BdPPbar~\cite{LHCb-PAPER-2013-038}.
No other two-body charmless baryonic $B$ decay modes have been observed.
Their experimental study requires large data samples,
presently only available at the LHC, as baryonic $B$ decays to two-body final states
are suppressed, with branching fractions typically one to two orders of magnitude lower
than similar baryonic decays to multibody final states.

Experimental input on the branching fractions of the \BuPLbar decay and other
suppressed baryonic decays provides valuable information on the dynamics of the decays
of $B$ mesons to baryonic final states.
The \BuPLbar decay mode is expected to be dominated by a $b\rightarrow s$ loop transition,
but tree-level (\Vub suppressed) and annihilation diagrams also contribute.
Various theoretical predictions for its branching fraction are available.
Calculations based on QCD sum rules~\cite{Chernyak1990137} predict a branching fraction
smaller than $3 \times 10^{-6}$ whereas a pole model~\cite{PhysRevD.66.014020}
and a recent study~\cite{PhysRevD.89.056003}, taking into account the \lhcb experimental
result on the $B^{0}\rightarrow p\bar{p}$ branching fraction~\cite{LHCb-PAPER-2013-038},
both predict a branching fraction around $2 \times 10^{-7}$.
The violation of partial conservation of the axial-vector current at the \gev scale
has been proposed as an alternative approach to the understanding of the data available
on two-body baryonic decays of $B$ and \Ds mesons~\cite{Hsiao:2014zza}.
It explains the \lhcb results on the \BPPbar decay modes~\cite{LHCb-PAPER-2013-038}
and predicts a branching fraction for the \BuPLbar decay of the order of $10^{-8}$.

The decay \BuPLbar has been searched for by the \cleo~\cite{Coan:1998hy}
and \belle~\cite{Tsai:2007pp} collaborations. The most
stringent experimental upper limit on the \BuPLbar branching fraction is
$3.2 \times 10^{-7}$ at 90\% confidence level,
determined by the \belle collaboration using 414\invfb of integrated luminosity
from \epem collisions.

This paper presents a search for the rare decay mode \BuPLbar
with the full $pp$ collision data sample collected in 2011 and 2012 by the \lhcb experiment.
The branching fraction is measured with respect to that of the topologically identical \BuKSPi decay
to suppress common systematic uncertainties.
The $\PLambda$ baryon is reconstructed in the \LPPi final state whereas the \KS meson
is reconstructed in its \KSPiPi final state.
The inclusion of charge-conjugate processes is implied throughout this paper.

%% file: detector.tex
\section{Detector and data sample}
\label{sec:detector}
The data sample analysed corresponds to an integrated luminosity of 1\invfb at a centre-of-mass energy
of 7\tev recorded in 2011 and 2\invfb at 8\tev recorded in 2012.
The \lhcb detector~\cite{Alves:2008zz,LHCb-DP-2014-002} is a single-arm forward
spectrometer covering the \mbox{pseudorapidity} range $2<\eta <5$,
designed for the study of particles containing \bquark or \cquark
quarks. The detector includes a high-precision tracking system
consisting of a silicon-strip vertex detector surrounding the $pp$
interaction region, a large-area silicon-strip detector located
upstream of a dipole magnet with a bending power of about
$4{\mathrm{\,Tm}}$, and three stations of silicon-strip detectors and straw
drift tubes placed downstream of the magnet.
The tracking system provides a measurement of momentum, \ptot, of charged particles with
a relative uncertainty that varies from 0.5\% at low momentum to 1.0\% at 200\gevc.
The minimum distance of a track to a primary vertex (PV), the impact parameter (IP), 
is measured with a resolution of $(15+29/\pt)\mum$,
where \pt is the component of the momentum transverse to the beam, in\,\gevc.
The different types of charged hadrons are distinguished using information
from two ring-imaging Cherenkov detectors.
Photons, electrons and hadrons are identified by a calorimeter system consisting of
scintillating-pad and preshower detectors, an electromagnetic
calorimeter and a hadronic calorimeter. Muons are identified by a
system composed of alternating layers of iron and multiwire
proportional chambers.

The decays of the \Vz\ hadrons, namely \LPPi\ and \KSPiPi, are
reconstructed in two different categories: the first consists of \Vz\ hadrons
that decay early enough for the daughter particles to be
reconstructed in the vertex detector, and the second contains
those that decay later such that track segments cannot be reconstructed in
the vertex detector. These categories are referred to as \emph{long}
and \emph{downstream}, respectively. The candidates in the long category have better
mass, momentum and vertex resolution than those in the downstream category.

Events are selected in a similar way for
both the \BuPLbar signal decay and the normalisation channel \BuKSPi.
The online event selection is performed by a trigger consisting of a hardware stage,
based on information from the calorimeter and muon systems, followed
by a software stage that performs a full event reconstruction, in which
all charged particles with $\pt > 500\,(300)\mevc$ are reconstructed for the
2011 (2012) data.  At the hardware trigger stage, events are required
to have a muon with high \pt or a hadron, photon or electron with high
transverse energy in the calorimeters.
The transverse energy threshold for hadrons is set at 3.5\gev.
Signal candidates may come from events where the hardware trigger was activated
either by signal particles or by other particles in the event.
The proportion of events triggered by other particles in the event is found to be very similar
between the signal and the normalisation decay modes in both long and downstream samples.
The software trigger requires a two- or
three-track secondary vertex with a significant displacement
from the primary $pp$ interaction vertices. At least one charged
particle must have $\pt > 1.7\gevc$ and be
inconsistent with originating from a PV.  A multivariate
algorithm~\cite{BBDT} is used for the identification of secondary
vertices consistent with the decay of a \bquark hadron
to a final state of two or more particles.

The efficiency of the software trigger selection on both decay
modes varied during the data-taking period.
During the 2011 data taking, downstream tracks were not reconstructed in the software
trigger. Such tracks were included in the trigger during the 2012
data taking and a further significant improvement in the algorithms was
implemented mid-year. Consequently, the data are
subdivided into three data-taking periods (2011, 2012a and 2012b)
in addition to the two \Vz reconstruction categories (long and downstream).
The 2012b sample has the highest trigger efficiency, especially in the
downstream category, and is also the largest data set, corresponding to
1.4\invfb of integrated luminosity.

Simulated data samples are used to study the response of the detector and to
investigate possible sources of background to the signal and the normalisation modes.
The $pp$ collisions are generated using \pythia~\cite{Sjostrand:2006za,*Sjostrand:2007gs} 
with a specific \lhcb configuration~\cite{LHCb-PROC-2010-056}.
Decays of hadronic particles
are described by \evtgen~\cite{Lange:2001uf}, in which final-state
radiation is generated using \photos~\cite{Golonka:2005pn}. The
interaction of the generated particles with the detector, and its response,
are implemented using the \geant toolkit~\cite{Allison:2006ve, *Agostinelli:2002hh}
as described in Ref.~\cite{LHCb-PROC-2011-006}.

%% file: selection.tex
\section{Sample selection and composition}
\label{sec:selection}
The selection consists of two stages, a preselection with high efficiency for the signal decays,
followed by a multivariate classifier.
The selection requirements of both signal and normalisation decays
exploit the characteristic topology and kinematic properties of two-body decays to
final states containing a \Vz hadron.
The \Bu candidates are reconstructed by combining, in a good-quality vertex,
a \Vz candidate with a charged particle hereafter referred to as the bachelor particle.
Both the \BuPLbar and the \BuKSPi decay chains are refitted~\cite{Hulsbergen:2005pu}
using the known \Lz or \KS mass~\cite{PDG2014}.
The resulting \Bu invariant-mass resolutions are improved and nearly identical
for the long and downstream \Vz candidates.
The long and downstream samples are thus merged after full selection,
thereby simplifying the extraction of the signal yields.

A minimum \pt requirement is imposed for all final-state particles.
The \Vz decay products must have a large IP with respect to all PVs;
hence a minimum \chisqip with respect to the PVs
is imposed on each decay product, where \chisqip is defined as the difference
between the vertex-fit \chisq of a PV reconstructed with and without the track in question.
The \Vz decay products are also required to form a good quality vertex.
The \Vz candidate is associated to the PV that gives the smallest \chisqip.
The selection favours long-lived \Vz decays
by requiring that the decay vertex and the associated PV are well separated.

The $\PLambda$ decay products must satisfy $|m(\proton \pi) - m_\PLambda| < 20 (15)\mevcc$
for downstream (long) candidates, where $m_\PLambda$ is the known $\PLambda$ mass~\cite{PDG2014}.
The corresponding criterion for the \KS decay products is $|m(\pi \pi) - m_\KS| < 30 (15)\mevcc$,
where $m_\KS$ is the known \KS mass~\cite{PDG2014}.

The \Bu candidate is required to have a small \chisqip with respect to the associated PV
as its reconstructed momentum vector should point to its production vertex.
This pointing condition of the \Bu candidate is further reinforced
by requiring that the angle between the \Bu candidate momentum vector and the line
connecting the associated PV and the \Bu decay vertex (\Bu direction angle) is close to zero.

To avoid selection biases, \PLbar candidates with invariant mass within
$64 \mevcc$ (approximately four times the mass resolution)
around the known \Bu mass are not examined until all
analysis choices are finalised. No such procedure is applied to the spectrum of the
well-known \BuKSPi decay.
The final selections of \PLbar and \KSPi candidates rely on
artificial neural networks~\cite{MLPs}, multilayer perceptrons~(MLPs),
as multivariate classifiers to separate signal from background;
the MLP implementation is provided by the TMVA toolkit~\cite{TMVA}.

Separate MLPs are employed for the \PLbar and the \KSPi selection.
The MLPs are trained with simulated samples to represent the signals
and with data from the high-mass sideband in the range 5350--6420\mevcc for the background,
to avoid partially reconstructed backgrounds.
For the well-known \KSPi spectra both low- and high-mass sidebands are used.
The training and selection is performed separately for each period of data taking
(the 2012a and 2012b samples are merged) and for downstream and long samples.
Optimisation biases are avoided by splitting each of these samples into three disjoint subsamples:
each MLP is trained on a different subsample in such a way that
events used to train one MLP are classified with another.
The response of the MLPs is uncorrelated with the mass of the \PLbar and \KSPi final states.
The MLP training relies on an accurate description of the distributions of the input
variables in simulated events.
The agreement between data and simulation is verified with kinematic distributions
from \BuKSPi decays, where the combinatorial background in the invariant mass spectrum
is statistically suppressed using the \splot technique~\cite{Pivk:2004ty}.
No significant deviations are found, giving confidence that the inputs to
the MLPs represent the data reliably. The variables used in the
MLP classifiers are properties of the \Bu candidate and of the
bachelor particle and \Vz daughters.  The input variables are the following:
the \chisq per degree of freedom of the kinematic fit of the decay chain; the \Bu decay
length, \chisqip and direction angle; the difference between the
$z$-positions of the \Bu and the \Vz decay vertices divided by its
uncertainty squared; the bachelor particle \pt; and the \pt of the \Vz decay products.
Extra variables are exploited in the selection of the long samples:
the \chisqip of the bachelor particle and of both \Vz decay products.

In addition to the MLP selection, particle identification (PID) requirements
are necessary to reject sources of background coming from $B$ decays.
A loose PID requirement is imposed on the \Vz daughters,
exploiting information from the ring-imaging Cherenkov detectors,
to remove background from \KS (\Lz) decays in the \PLbar (\KSPi) samples.
The PID selection on the bachelor particle is optimised together
with the MLP selection as follows.
The figure of merit $\epsilon_{\rm sig}/(a/2 + \sqrt{B_{\rm exp}})$
suggested in Ref.~\cite{Punzi:2003bu} is used to determine the optimal
MLP and PID requirements for each \BuPLbar subsample separately,
where $\epsilon_{\rm sig}$ represents the combined MLP and PID selection efficiency.
The term $a = 3$ quantifies the target level of significance in units of standard deviations.
The expected number of background candidates, $B_{\rm exp}$,
within the (initially excluded) signal region
is estimated by extrapolating the result of a fit to the invariant mass
distribution of the data sidebands.
A standard significance $S/\sqrt{S+B}$ is used to optimise the selection
of the \BuKSPi candidates, where $B$ is the number of background candidates
and $S$ the number of signal candidates in the invariant mass range $5000-5600\mevcc$.
The presence of the cross-feed background \BuKSK is taken into account.
The fraction of events with more than one selected candidate is negligible;
all candidates are kept.

Efficiencies are determined for each data-taking period and each \Vz reconstruction category,
and subsequently combined accounting for the mixture of these subsamples in data.
The efficiency of the MLP selection is determined from simulation.
Large data control samples of $\Dz \to \Km \pip$, \LPPi  and \LcPKPi decays
are employed~\cite{LHCb-DP-2012-003} to determine the efficiency of the PID requirements.
All other selection efficiencies, \ie\ trigger, reconstruction and preselection efficiencies,
are determined from simulation.
The overall selection efficiencies of this analysis are of order $10^{-4}$.
The expected yield of the control mode \BuKSPi, calculated from the product of the
integrated luminosity, the $\bquark\bquarkbar$ cross-section, the $\bquark$ hadronisation probability,
the \BuKSPi visible branching fraction and the total selection efficiency,
agrees with the yield obtained from the fit to the data at the level of 1.4 standard deviations.

Possible sources of non-combinatorial background to the \PLbar and \KSPi spectra are investigated
using extensive simulation samples. These sources include partially reconstructed backgrounds
in which one or more particles from the decay of a $b$ hadron are not associated with the
signal candidate, and $b$-hadron decays where one or more decay products are misidentified,
such as decays with \KS mesons misidentified as \Lz baryons in the \PLbar spectrum.
The peaking background from \BuPPbarPi decays in the \PLbar spectrum is found to be
insignificant after the MLP selection.
The currently unobserved \BuPSbar decay is treated as a source of systematic uncertainty.
The ensemble of specific backgrounds does not peak
in the signal region but rather contributes a smooth \PLbar mass spectrum,
which is indistinguishable from the dominant combinatorial background.

%% file: fits.tex
\section{Signal yield determination}
\label{sec:fits}
The yields of the signal and background candidates in both the signal and normalisation samples
are determined, after the full selection, using unbinned extended maximum likelihood fits
to the invariant mass spectra.
The signal lineshapes are found to be compatible between the data-taking periods and between
the long and downstream categories, so all subsamples are merged together into a single spectrum.

The probability density functions (PDFs) of $B$-meson signals have
asymmetric tails that result from a combination of detector-related effects
and effects of final-state radiation. The signal mass distributions are verified
in simulation to be modelled accurately by the sum of two Crystal Ball (CB)
functions~\cite{Skwarnicki:1986xj} describing the high- and low-mass asymmetric tails.
The peak values and the core widths of the two CB components are set to be the same.

\begin{figure}[bt]
  \begin{center}
    \includegraphics[width=0.8\linewidth]{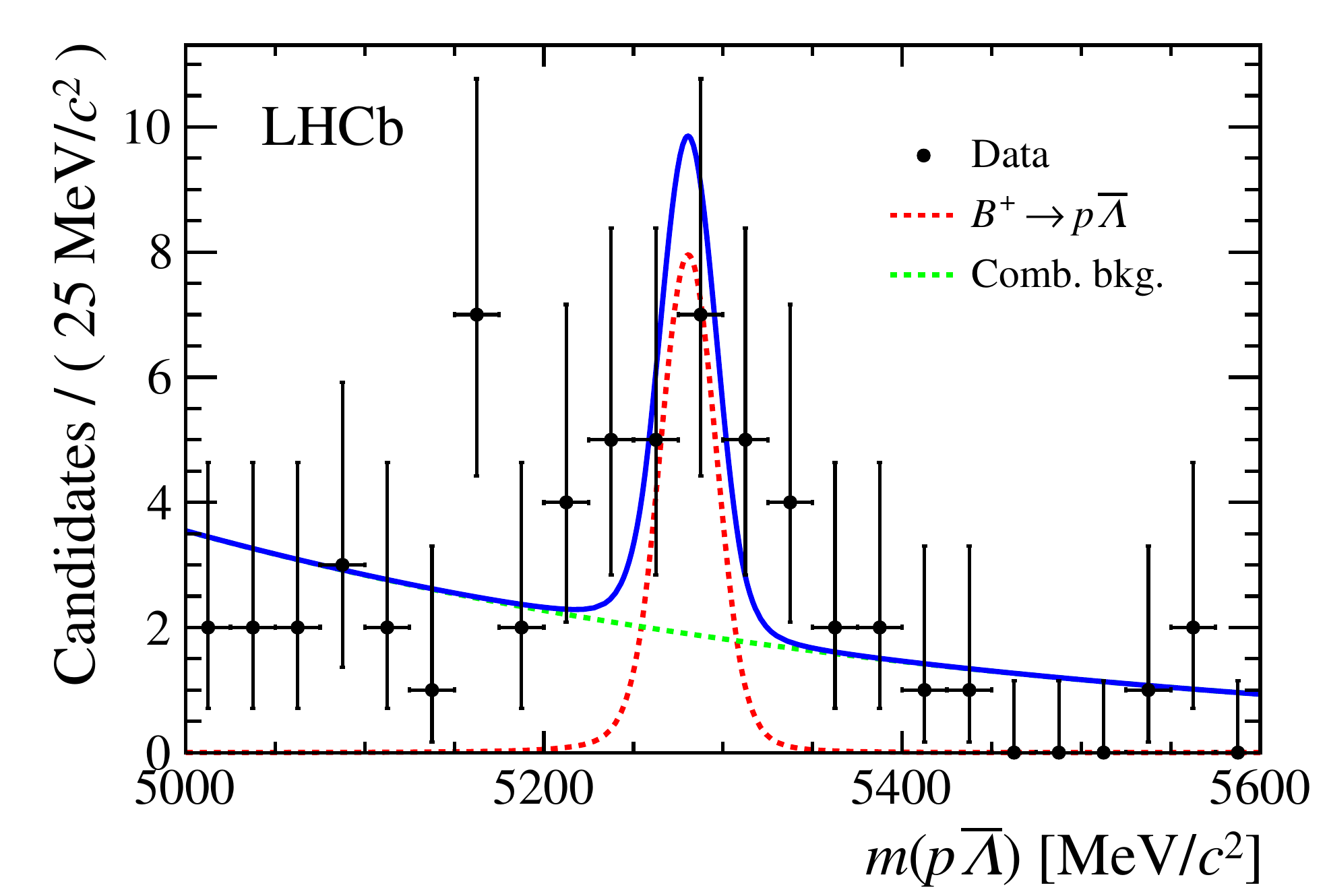}
    \vspace*{-0.7cm}
  \end{center}
  \caption{
    Invariant mass distribution of \PLbar candidates after full selection.
    The result of the fit to the data (blue, solid) is shown together with each fit model component,
    namely the \BuPLbar signal and the combinatorial background.}
  \label{fig:fit_PLbar_spectrum}
\end{figure}

The \PLbar spectrum comprises the \BuPLbar signal and combinatorial background.
Contamination from partially reconstructed backgrounds,
with or without misidentified particles, is treated as a source of systematic uncertainty.
The peak position and tail parameters of the \BuPLbar CB components are fixed to the values
obtained from simulation. The core width parameter, also fixed, is obtained by multiplying the value
from simulation by a scaling factor to account for differences in the resolution between data and simulation.
This factor, determined from the \BuKSPi data and simulation samples,
is compatible with unity ($1.01\pm0.06$) and gives a width of approximately 16\mevcc.
The invariant mass distribution of the combinatorial background is described by an exponential function,
with the slope parameter determined from the fit.

The fit to the \PLbar invariant mass distribution, presented in Fig.~\ref{fig:fit_PLbar_spectrum},
determines three parameters: two yields and the slope of the combinatorial background model.
An excess of \BuPLbar candidates with respect to background expectations is found,
corresponding to a signal yield of $N(\BuPLbar) = 13.0 ^{+5.1}_{-4.3}$,
where the uncertainties, obtained from a profile likelihood scan, are statistical only.

The statistical significance of the \BuPLbar signal is determined with a
large set of samples simulated assuming the presence of background only.
For each simulated sample, a number of events distributed according to the
exponential model of the background is drawn from a Poisson distribution
with mean equal to the number of observed background events.
For each sample, the log-likelihood ratio
$2\ln(L_{S+B}/L_{B})$ is computed, where $L_{S+B}$ and
$L_{B}$ are the likelihoods from the full fit and from the fit
without the signal component, respectively.
The fraction of samples that yield log-likelihood ratios larger than
the ratio observed in data is $3.4 \times 10^{-5}$,
which corresponds to a statistical significance of 4.1 standard deviations.
Inclusion of the 6.7\% systematic uncertainty affecting the signal yield
gives only a marginal change in the signal significance.

\begin{figure}[tb]
  \begin{center}
    \includegraphics[width=0.8\linewidth]{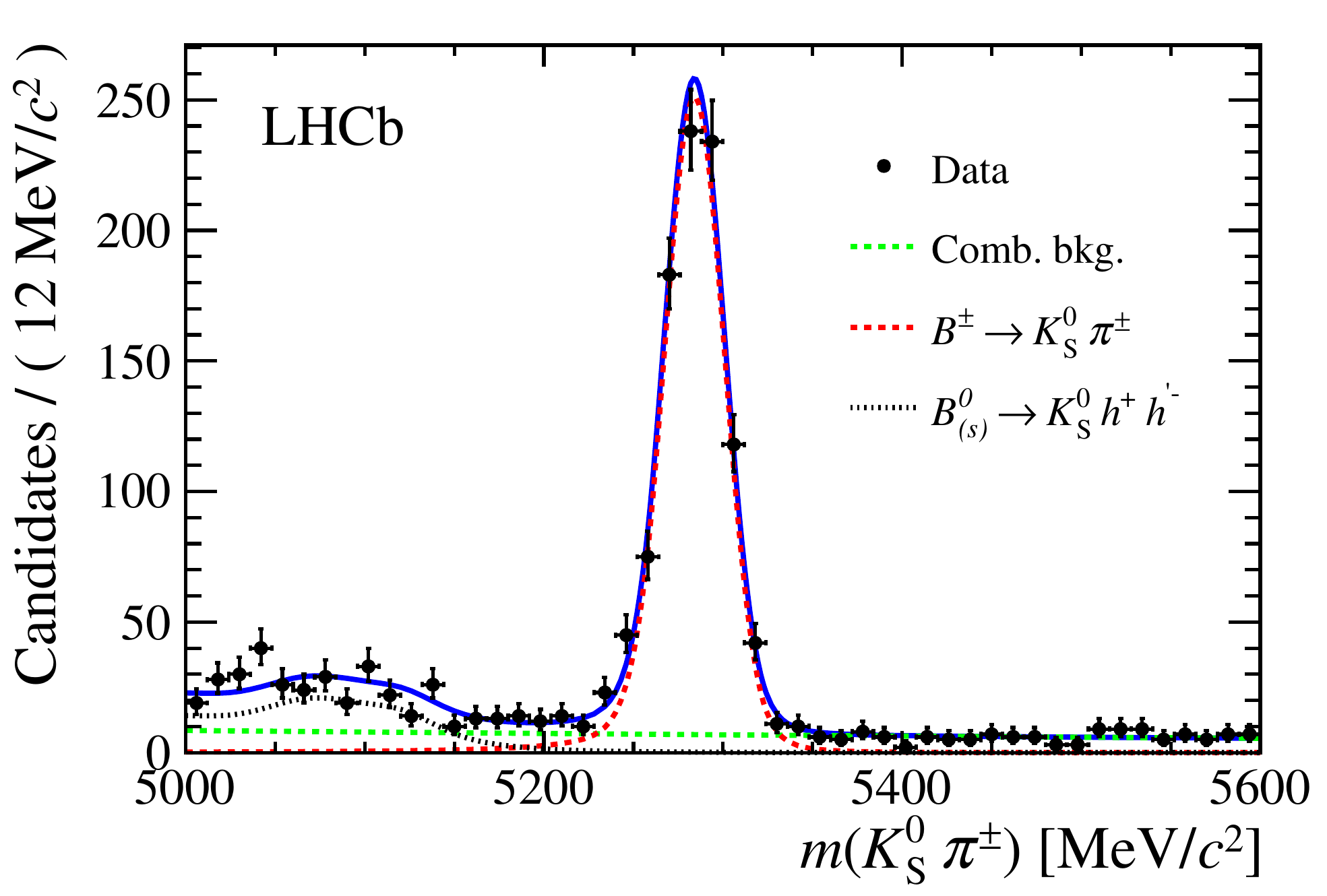}
    \vspace*{-0.7cm}
  \end{center}
  \caption{
    Invariant mass distribution of \KSPi candidates after full selection.
    The result of the fit to the data (blue, solid) is shown together with each fit model component,
    namely the \BuKSPi signal, the \BKSHH partially reconstructed background,
    and the combinatorial background.
    The vanishingly small \BuKSK misidentified cross-feed is not displayed.}
  \label{fig:fit_KSPi_spectrum}
\end{figure}

The \KSPi mass spectrum of the normalisation decay is described as the sum of components
accounting for the \BuKSPi signal, the \BuKSK misidentified background,
backgrounds from partially reconstructed \BKSHH decays (\mbox{$h^{(\prime)} = \pion, \kaon$}),
and combinatorial background.
Any contamination from other decays is treated as a source of systematic uncertainty.

The \BuKSH CB tail parameters and the relative normalisation of the two CB functions
are fixed to the values obtained from simulation.
The mean and the width (approximately 17\mevcc) of the \BuKSPi peak are allowed to vary
in the fit to the data, whilst they are fixed for the very small \BuKSK peak contribution.
The mean of the \BuKSK peak, around 5240\mevcc, is fixed using the mass difference between the
\BuKSK and \BuKSPi peaks obtained from simulation.
The \BuKSK yield is Gaussian constrained using the \BuKSPi yield, taking into account the differences
in branching fraction and selection efficiency.

The partially reconstructed backgrounds that populate the lower-mass sideband are assumed to arise
from the \mbox{\BKSHH} decay modes with the largest branching fractions,
namely \BdKSPiPi, \BdKSKK, \BsKSPiPi and \BsKSKPi~\cite{LHCb-PAPER-2013-042}.
Only \BdKSPiPi, \BsKSPiPi and \BsKSKPi are considered given that \mbox{\BdKSKK}
is further suppressed because of a low kaon-to-pion misidentification probability.
The overall shape of the \BKSHH decay modes in the \KSPi mass spectrum is
obtained from simulation accounting for the relative yields
related to different $B$-meson fragmentation probabilities, selection efficiencies and branching fractions.
The invariant mass distribution of the combinatorial background is described by
an exponential function, with the slope parameter determined from the mass fit.

The resulting spectrum shows a prominent \BuKSPi peak above little combinatorial
and partially reconstructed background.
The fit to the \KSPi spectrum, presented in Fig.~\ref{fig:fit_KSPi_spectrum},
determines seven parameters: three shape parameters and four yields.
The signal yield obtained is $N(\BuKSPi) = 930 \pm 34$,
where the uncertainty is statistical only.

%% file: systematics.tex
\section{Systematic uncertainties}
\label{sec:systematics}
The systematic uncertainties are reduced by performing the branching fraction
measurement relative to a decay mode topologically identical to the decay of interest.
Uncertainties arise from imperfect knowledge of the selection efficiencies,
systematic uncertainties on the fitted yields, and uncertainties on the branching fractions
of decays involved in the calculation of the \BuPLbar branching fraction.
The systematic uncertainties assigned to the measurement of the \BuPLbar branching fraction
are summarised in Table~\ref{tab:syst_summary}.

\begin{table}[tb]
  \caption{\small{Summary of systematic uncertainties relative to the
           measured \BuPLbar branching fraction.
           The contributions are split into those that come from the signal decay
           and those that come from the normalisation decay.
           The total corresponds to the sum of all contributions added in quadrature.}}
  \begin{center}
  \vspace{\baselineskip}
  \begin{tabular}{l|cc}
    Source                              & \multicolumn{2}{c}{Value [\%]}\\
                                        & \BuPLbar & \BuKSPi\\
    \hline
    $\BF(\BuKSPi)$                      & -        & 3.2\\
    Trigger efficiencies ratio          & 3.5      & -  \\
    Selection efficiencies ratio        & 2.2      & -\\
    PID uncertainties                   & 1.2      & 3.5\\
    Tracking efficiencies ratio         & 6.0      & - \\
    Yields from mass fits               & 6.7      & 3.0\\
    Simulation statistics               & 1.7      & 3.3  \\
    \hline
    Total                               & 10.1     & 6.5  \\
  \end{tabular}
  \end{center}
  \label{tab:syst_summary}
\end{table}

The uncertainty on the branching fraction of the normalisation channel,
\mbox{$\BF(\BuKSPi)=(11.895\pm0.375)\times 10^{-6}$~\cite{HFAG}}
(assuming that half of the \Kz mesons decay as a \KS), is taken as a systematic uncertainty.
The uncertainties on the branching fractions $\BF(\LPPi) = (63.9\pm0.5)\%$ and
$\BF(\KSPiPi) = (69.20\pm0.05)\%$ are accounted for, but omitted from the table
as they are negligible compared to all other sources of systematic uncertainty.

The determination of the selection efficiencies entails several sources of
systematic uncertainty.
A systematic uncertainty is assigned to take into account possible
differences in the trigger efficiencies between data and simulation,
following the procedure and studies described in Ref.~\cite{LHCb-PAPER-2016-002}.
The \BuKSPi mode is used as a proxy for the assessment of the systematic uncertainties
related to the MLP selection. Distributions for the \BuKSPi MLP input variables
are obtained from data using the \splot technique. The distributions from simulation
showing the largest discrepancies are weighted to match those of the data.
The selection efficiencies are recalculated with the same MLPs,
but using the weighted distributions, to derive the variations in efficiency,
and hence the systematic uncertainty on the selection.
The uncertainty associated with the imperfect knowledge of PID selection efficiencies
is assessed varying the binning of the PID control samples in track momentum and
pseudorapidity, and also accounting for a dependence of the efficiency
on the event track multiplicity after weighting the distribution of the latter
to match that of the data. The two uncertainties are combined in quadrature.

The signal decay has two baryons in the final state whilst only mesons are present
in the final state of the normalisation channel.
Tracking efficiency uncertainties do not cancel fully in this instance.
The degree to which the simulation describes the hadronic interactions
with the material is less accurate for baryons than it is for mesons.
A systematic uncertainty of 4\% per proton is estimated
whereas the corresponding uncertainty is 1.5\% for pions and kaons~\cite{LHCb-DP-2013-002}.
A non-negligible systematic uncertainty on the tracking efficiencies as calculated from
simulation, including correlations, results from these sources of uncertainty.

Systematic uncertainties on the fit yields arise from potential mismodelling
of the fit components and from the uncertainties
on the values of the parameters fixed in the fits.
They are investigated using data and by studying a large number of simulated data samples,
with parameters varying within their estimated uncertainties.
Changing the combinatorial background model to a linear shape decreases the signal yield
by 4.6\%, with no significant effect on the signal significance and the final result.
Possible contamination from the unobserved decay \BuPSbar is studied by adding such a component.
The fitted \BuPSbar yield is found to be compatible with zero,
and the shift in the \BuPLbar yield with respect to the nominal yield is taken as a systematic uncertainty.

The finite size of the simulation samples used in the analysis further contributes
as a source of systematic uncertainty.
The total systematic uncertainty on the \BuPLbar branching fraction is given by the sum
of all contributions added in quadrature, amounting to 12.0\%.

%% file: conclusion.tex
\section{Results and conclusion}
\label{sec:results_conclusion}
The \BuPLbar branching fraction is determined relative to that of the
\BuKSPi normalisation channel according to
\begin{equation}
\BF(\BuPLbar) = \frac{N(\BuPLbar)}{N(\BuKSPi)}\,
                \frac{\epsilon_{\BuKSPi}}{\epsilon_{\BuPLbar}}\,
                \frac{\BF(\KSPiPi)}{\BF(\LPPi)}\,
                \BF(\BuKSPi)\,,\nonumber
  \label{eq:yield_norm}
\end{equation}
where $N$ represent the yields determined from the
mass fits and $\epsilon$ are the selection efficiencies.
It is measured to be
\begin{equation}
\BF(\BuPLbar) = ( 2.4 \,^{+1.0}_{-0.8} \pm 0.3 ) \times 10^{-7} \,,\nonumber
\end{equation}
where the first uncertainty is statistical and the second systematic.

In summary, a search is reported for the rare two-body charmless
baryonic decay \BuPLbar using a $pp$ collision data sample collected by the \lhcb experiment
at centre-of-mass energies of 7 and 8~\tev,
corresponding to an integrated luminosity of 3\invfb.
An excess of \BuPLbar candidates with respect to background
expectations is found with a statistical significance of 4.1 standard deviations.
This is the first evidence for this decay process.

The measured branching fraction is compatible with the
theoretical predictions in Refs.~\cite{PhysRevD.66.014020,PhysRevD.89.056003}
but is in tension with calculations based on QCD sum rules~\cite{Chernyak1990137}
and calculations based on factorisation with the hypothesis of the violation of
partial conservation of the axial-vector current at the \gev scale~\cite{Hsiao:2014zza}.
It helps shed light on an area of hadronic physics in which experimental input
is needed, namely the study of the mechanisms responsible for decays
of $B$ mesons to baryonic final states.

%% file: acknowledgements.tex
\section*{Acknowledgements}
\noindent We express our gratitude to our colleagues in the CERN
accelerator departments for the excellent performance of the LHC. We
thank the technical and administrative staff at the LHCb
institutes. We acknowledge support from CERN and from the national
agencies: CAPES, CNPq, FAPERJ and FINEP (Brazil); NSFC (China);
CNRS/IN2P3 (France); BMBF, DFG and MPG (Germany); INFN (Italy); 
FOM and NWO (The Netherlands); MNiSW and NCN (Poland); MEN/IFA (Romania); 
MinES and FASO (Russia); MinECo (Spain); SNSF and SER (Switzerland); 
NASU (Ukraine); STFC (United Kingdom); NSF (USA).
We acknowledge the computing resources that are provided by CERN, IN2P3 (France), KIT and DESY (Germany), INFN (Italy), SURF (The Netherlands), PIC (Spain), GridPP (United Kingdom), RRCKI and Yandex LLC (Russia), CSCS (Switzerland), IFIN-HH (Romania), CBPF (Brazil), PL-GRID (Poland) and OSC (USA). We are indebted to the communities behind the multiple open 
source software packages on which we depend.
Individual groups or members have received support from AvH Foundation (Germany),
EPLANET, Marie Sk\l{}odowska-Curie Actions and ERC (European Union), 
Conseil G\'{e}n\'{e}ral de Haute-Savoie, Labex ENIGMASS and OCEVU, 
R\'{e}gion Auvergne (France), RFBR and Yandex LLC (Russia), GVA, XuntaGal and GENCAT (Spain), Herchel Smith Fund, The Royal Society, Royal Commission for the Exhibition of 1851 and the Leverhulme Trust (United Kingdom).

%% file: LHCb_Authorship_flat_20-Sep-2016.tex
\centerline{\large\bf LHCb collaboration}
\begin{flushleft}
\small
R.~Aaij$^{40}$,
B.~Adeva$^{39}$,
M.~Adinolfi$^{48}$,
Z.~Ajaltouni$^{5}$,
S.~Akar$^{6}$,
J.~Albrecht$^{10}$,
F.~Alessio$^{40}$,
M.~Alexander$^{53}$,
S.~Ali$^{43}$,
G.~Alkhazov$^{31}$
% ...

{\footnotesize \it
% ...
$ ^{5}$Clermont Universit{\'e}, Universit{\'e} Blaise Pascal, CNRS/IN2P3, LPC, Clermont-Ferrand, France\\
$ ^{6}$CPPM, Aix-Marseille Universit{\'e}, CNRS/IN2P3, Marseille, France\\
% ...
$ ^{10}$Fakult{\"a}t Physik, Technische Universit{\"a}t Dortmund, Dortmund, Germany\\
% ...
$ ^{31}$Petersburg Nuclear Physics Institute (PNPI), Gatchina, Russia\\
% ...
$ ^{39}$Universidad de Santiago de Compostela, Santiago de Compostela, Spain\\
$ ^{40}$European Organization for Nuclear Research (CERN), Geneva, Switzerland\\
% ...
$ ^{43}$Nikhef National Institute for Subatomic Physics, Amsterdam, The Netherlands\\
% ...
$ ^{48}$H.H. Wills Physics Laboratory, University of Bristol, Bristol, United Kingdom\\
% ...
$ ^{53}$School of Physics and Astronomy, University of Glasgow, Glasgow, United Kingdom\\
% ...
}
\end{flushleft}